# Allocation of control and data channels for Large-Scale Wireless Sensor Networks

Jamila BEN SLIMANE[1,2], Ye-Qiong SONG[2], Anis KOUBAA[3,4], Mounir FRIKHA[1]

[1] Sup'Com-MEDIATRON, City of Communication Technologies, 2083 Ariana, Tunisia
[2] LORIA and INPL, Campus Scientifique, BP 239 54506 Vandoeuvre-les-Nancy, France
[3] IPP-HURRAY! Research Group, Polytechnic Institute of Porto, Rua António Bernardino de Almeida, 431, 4200-072 Porto, Portugal
[4] Al-Imam Muhammad ibn Saud University, Computer Science Dept.,11681 Riyadh, Saudi Arabia
Email: jamilabs07@yahoo.fr, Song@loria.fr, akoubaa@dei.isep.ipp.pt, m.frikha@supcom.rnu.tn

## I – Introduction

Both IEEE 802.15.4 and 802.15.4a standards allow for dynamic channel allocation and use of multiple channels available at their physical layers but its MAC protocols are designed only for single channel. Also, sensor's transceivers such as CC2420 provide multiple channels and as shown in [1], [2] and [3] channel switch latency of CC2420 transceiver is just about 200µs.
In order to enhance both energy efficiency and to shorten end to end delay, we propose, in this report, a spectrum-efficient frequency allocation schemes that are able to statically assign control channels and dynamically reuse data channels for Personal Area Networks (PANs) inside a Large-Scale WSN based on UWB technology.

## II - System Model

### II - 1 - Network Topology

In order to deploy a dense network supporting a considerable number of nodes, we proposed in [4] a three-tiered network to represent the global network, using UWB sensors in the first and second network levels. The choice of the UWB technology is done to benefit from its extreme low transmit power minimizing interference, high data rate allowing real time and high data rate applications and location capacity allowing mobility management and node identification. For the third tier, we propose Wifi network to benefit from its high data rate, large coverage and security aspect. What we aim is an application in hospital where the global network represents WHSN (Wireless Hospital Sensor Network). Fig.1 shows all network layers composing the WHSN.

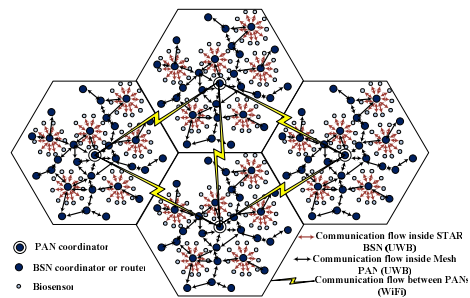

Fig.1 WHSN architecture

The lowest level represents the Body Sensor Network (BSN). We can model an elementary BSN by a star network composed of one coordinator and a set of biosensors that ensure physiological measurements and medical monitoring of patient. To improve patient's network performance in a dense hospital environment, we propose overlaying the network of BSNs with a second upper level network. The hexagon cell represents the Personal Area Network (PAN) or the second network level. As shown in Fig.1, the network is represented by a cell of sensors organized in mesh topology including one PAN coordinator, several mobile BSNs coordinators (one coordinator per BSN) and several routers. For an efficient solution for channel allocation and mobility management in WHSNs, that cellular architecture, based on UWB/Wifi technologies, is chosen to the third level to have at the end a three-tier hierarchical cellular network.
The detailed description of the network architecture is out of scoop of this report, so for more details, one can refer to [4]. In this paper we are interested in UWB spectrum allocation problem at WHSN's second level (PAN). The problem of frequency allocation for WSNs is different from that treated in traditional cellular network such

GSM although we propose hexagonal cellular representation for the global network seen that each network have its proper specificity and requirements.

Let us assume the general case of a network composed by of $N_c$ PANs or hexagonal cells uniformly distributed as shown in Fig.2. The ideal case of a hexagonal model is chosen to ensure the totality coverage of the network. Although in practice the coverage zone of a sensor device is not an hexagon or a perfect circle, there are procedures and mechanisms [5] that ensure the adjustments of the model during network deployment by means of experimental test of measurements.

Let $H$ be a Cartesian coordinate system with $C_{0,0}$ as origin point with coordinates $(x_0, y_0)$, X for abscissa and Y for ordinate. We represent an arbitrary cell center as $C_{i,j}$ by its coordinates $(x_i, y_j)$ given by:

$$x_i = x_0 + i \times \left(\frac{3R}{2}\right), i \in [-N; N] \quad (1)$$

$$y_j = y_0 + j \times \left(\frac{\sqrt{3}R}{2}\right), j \in [-N; N] \quad (2)$$

$$(i+j) \bmod 2 = 0 \quad (3)$$

$$Card(\{(x_i, y_j) / (i,j) \in [-N; N]^2\}) = N_c \quad (4)$$

Let $C$ the set of all $C_{i,j}$ with coordinates $(x_i, y_j)$ verifying (1) to (4) as shown in Fig.2.

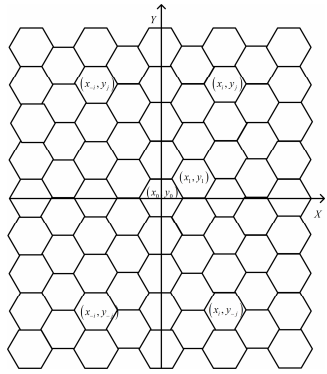

Fig.2 General case of a network of $N_c$ PANs

In the following sections only we are interested in the problem of UWB-channels sharing between PANs, seen that the problem of Wifi-channels sharing within a mesh network is already treated in [8] and [9].

## II - 2 - IEEE 802.15.4a IR-UWB SPECTRUM RESOURCE

IEEE 802.15.4a IR UWB complaint devices can operate in three independent bands: (1) the sub-gigahertz band (250-750 MHz), (2) the low band (3.1-5 GHz) and (3) the high band (6-10.6 GHz). Fig.3 gives the center frequencies and bandwidths of the admissible bands, as well as the regulatory domains in which they are admissible.

As shown in the table 39d given in [6], we dispose of 16 physical frequency channels associated with 8 sequence codes to have in total 32 logical channels.

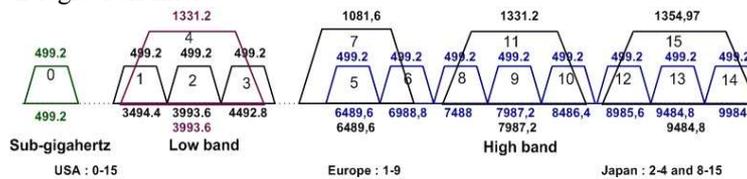

Fig.3 IEEE 802.15.4a UWB plan bands

According to table 39d given in [6] and Fig.3, neither overlapping channels nor adjacent channels share same sequence code. Consequently, overlapping channels don't represent co-channels since its sequence codes are different, in this case, the simultaneous use (in close space) of two overlapping channels don't produce co-channel interference. Also, adjacent logical channels don't interfere since its sequence codes are different.

Let us assume that $N_{tch}$ represents the set of total available logical channels. Conforming to worldwide UWB regulation $Card(N_{tch})$ is equal to 32, 18 and 22 for respectively US, Europe and Japan region.

According to radio transceiver characteristics, channel switch latency does not exceed 200μs. Although we can assume that during one duty cycle the additional delay introduced by switching radio channels is not significant, but an efficient channel-switch protocol must be proposed to avoid unnecessary channel switches that can degrade the network performance. To switch from a channel to another we need just to firstly programme the set of available frequency channels at the level of a specific register (e.g. FSCTRL.FREQ for CC2420 transceiver) then set this register to the adequate value to select demanded channel.

## III - STATIC CHANNELS ALLOCATION

### III - 1 - ALLOCATION OF CONTROL CHANNELS

*III - 1 - a - Case of network composed of 12 cells*

To avoid control channel congestion, we propose a static allocation of an optimal number of control channels. We assign one control channel to each PAN to persistently cover its cell from control traffic. We note that the overlapping channels (4, 7, 11 and 15) are more suitable to ensure the coverage of such traffic since they are characterized by its high bandwidth [6] allowing higher transmit power permitting an extended range compared to non-overlapping channels.

***Notations.***
- $N_c$ : The total number of cells,
- $R$ : Radius of a cell,
- $D_{N_c x N_c}$ : Distance matrix, distance separating each couple of cell centres,
- $N_{cch}$ : Set of available control channels that represents a sub set of total control channel set $N_{tcch}$.

$$N_{cch} \subseteq N_{tcch} = \begin{Bmatrix} (4, SC_7), (7, SC_7), (11, SC_7), (15, SC_7), \\ (4, SC_8), (7, SC_8), (11, SC_8), (15, SC_8) \end{Bmatrix}$$

The radius of different cells is the same and equals to the PAN coordinator coverage zone that we assume be circular with radius $R$ and all PAN members transmit with the same power transmit $P_0$. The choice of $P_0$ is done by taking into account the following equation.

$$P_{Rx} = P_0 + Pl(R) / P_{Rx} - Link\_margin = Rx\_sensitivity \qquad (5)$$

Pathloss expression is given in the 802.15.4a standard [6].

So, in this case all PAN's members precisely PAN members located at or near the cell border can hear their PAN coordinator control traffic (beacon frame,….), and it can be heard by their PAN coordinator. We can formulate this problem as 2-hop coloring problem, in which repetition of colors occurs only if the nodes belonging to different PANs are separated by more than 2 hops.

Consequently, as shown in Fig.4 the minimum distance of frequency reuse must be strictly bigger than 2 hops or distance $R_c$ (worst case). Let $D_{min}$ represents the minimal distance of frequency reuse referring to cells centers or positions of PANs coordinators:

$$D_{min} > R_c \qquad (6)$$

With $R_c$ is given by:

$$R_c = 2R \qquad (7)$$

From (6) and (7) and according to network hexagonal representation, the shortest distance frequency reuse at cells centers will be:

$$D_{min} = 4\left[R(\frac{\sqrt{3}}{2})\right] = (2\sqrt{3})R \qquad (8)$$

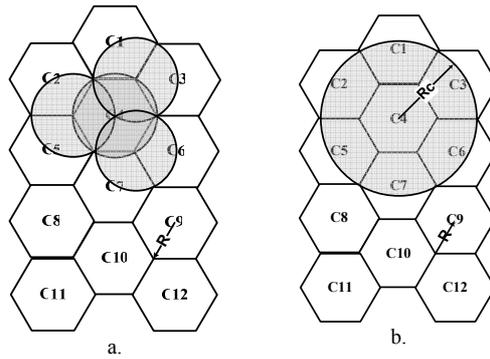

Fig.4 Radio coverage limit of a logical control channel

In this part, we are interested to find the minimal or optimal number of logical control channels $N_{cch-opt}$ ensuring a complete network coverage taking into account frequency reuse. This problem can be modelled as graph coloring problem "vertex coloring". As shown in Fig.6 we can represent our network as a $G(V,E)$ graph, where:
- Each cell center represents a vertex: $V$.
- Distance separating two cells centers that is shorter than $D_{min}$ represents edge: $E$.

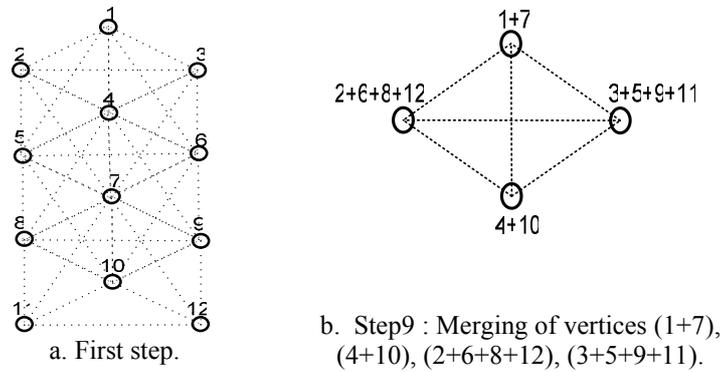

a. First step.  
b. Step9 : Merging of vertices (1+7), (4+10), (2+6+8+12), (3+5+9+11).

Fig.5 Logical control channel allocation graph

According to Fig.4 no two cells near than $D_{min}$ share the same control channel. For that we can call for one of optimal coloring algorithms such as Zykov's algorithm, branch and bound method, etc. The application of Zykov's algorithm to previous graph produces nine sub graphs.

In step 9, we are left with a complete graph ("A complete graph with n vertices obviously requires n colors" [7]). The optimal solution is given by the complete graph with 4 vertices in which vertices 1 and 7 are allocated to the first color, 2, 6, 8 and 12 to the second color 3, 5, 9 and 11 to the third color and vertices 4 and 10 to the fourth. We note that to cover the entire network by control traffic without suffering from co-channel interference, we just need of 4 different channel frequencies (See Fig.6 a and Fig.6 b)

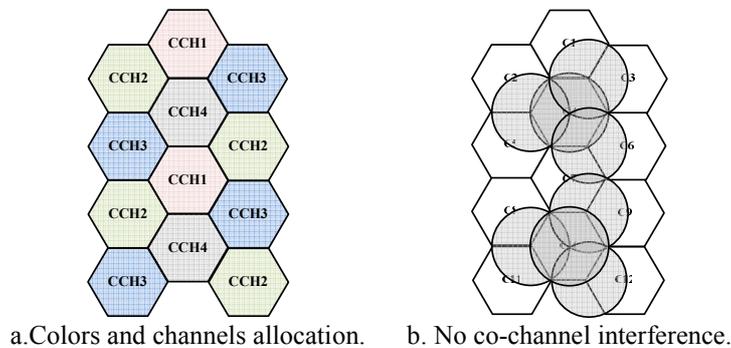

a. Colors and channels allocation.    b. No co-channel interference.

Fig.6 Logical control channel allocation

*III - 1 - b - General case*

**Theorem1**

Given a WSN composed by $N_c$ cells of radius $R$, to totally cover it by control traffic with one control channel per cell without suffering from co-channel interference, we need at most of 4 different channel frequencies.

**Proof1**

We can distinguish the following cases:
- Case $N_c = 1$:

  The number of needed control channels is equal to 1,
- Case Number of adjacent cells = 2:

  Let $C_{i,j}$ and $C_{m,n}$ be two adjacent cells (e.g. $d|C_{i,j}, C_{m,n}| < D_{min}$) $\Rightarrow$ the number of needed control channels is equal to 2,
- Case Number of adjacent cells = 3:

  Let $C_{i,j}, C_{m,n}$ and $C_{k,l}$ be three adjacent cells (e.g. $d|C_{i,j}, C_{m,n}| = d|C_{i,j}, C_{k,l}| = d|C_{k,l}, C_{m,n}| < D_{min}$) $\Rightarrow$ the number of needed control channels is equal to 3,
- Case Number of adjacent cells $\geq 4$ :

According to the example of 12 cells network, the shortest distance frequency reuse $D_{min}$ for control channels is equal to $(2\sqrt{3})R$. So to use the same frequency for a given couple of cells $(C_{i,j}, C_{m,n}) \subset C$, the distance separating the two cells must be equal to or bigger than $D_{min}$

$$d|C_{i,j}, C_{m,n}| \geq D_{min}$$

$$\sqrt{(x_i - x_m)^2 + (y_j - y_n)^2} \geq (2\sqrt{3})R$$

$$(x_i - x_m)^2 + (y_j - y_n)^2 \geq 12R^2$$

$$\left(\left[x_0 + i \times \left(\frac{3R}{2}\right)\right] - \left[x_0 + m \times \left(\frac{3R}{2}\right)\right]\right)^2 + \left(\left[y_0 + j \times \left(\frac{\sqrt{3}R}{2}\right)\right] - \left[y_0 + n \times \left(\frac{\sqrt{3}R}{2}\right)\right]\right)^2 \geq 12R^2 \quad (9)$$

$$\left(\left(\frac{3R}{2}\right) \times (i - m)\right)^2 + \left(\left(\frac{\sqrt{3}R}{2}\right) \times (j - n)\right)^2 \geq 12R^2$$

$$3 \times (i - m)^2 + (j - n)^2 \geq 16$$

Let sub sets $E_{ij}$, $F_{ij}$ and $G_{ij}$ given as follows:

$$E_{ij} = \begin{cases} C_{m,n} \in C / \\ 3 \times (i-m)^2 + (j-n)^2 > 16 \text{ and} \\ \text{fixed } C_{ij} \in C \end{cases}, \quad F_{ij} = \begin{cases} C_{m,n} \in C / \\ 3 \times (i-m)^2 + (j-n)^2 = 16 \text{ and} \\ \text{fixed } C_{ij} \in C \end{cases} \text{ and } G_{ij} = C - \{E_{ij} \cup F_{ij}\}$$

You can rewrite $F_{ij}$ as:

$$3 \times (i-m)^2 + (j-n)^2 = 16 \Rightarrow \begin{cases} (i-m)^2 = 4, (j-n)^2 = 4 \Rightarrow |i-m| = |j-n| = 2, \\ (i-m)^2 = 0, (j-n)^2 = 16 \Rightarrow i = m, |j-n| = 4 \end{cases} \quad (10)$$

Without loss of generality we can resolve (10) by considering simple case of $F_{00}$ then we deduce a general solution for $\forall F_{ij} \subset C$.

According to the definition of $F_{ij}$, $C_{m,n} \subset F_{00}$:

$$C_{m,n} \in \begin{Bmatrix} C_{04}, C_{22}, C_{2-2} \\ C_{0-4}, C_{-2-2}, C_{-22} \end{Bmatrix}$$

So, as shown in Fig.8, we can deduct $G_{00}$ from $F_{00}$.

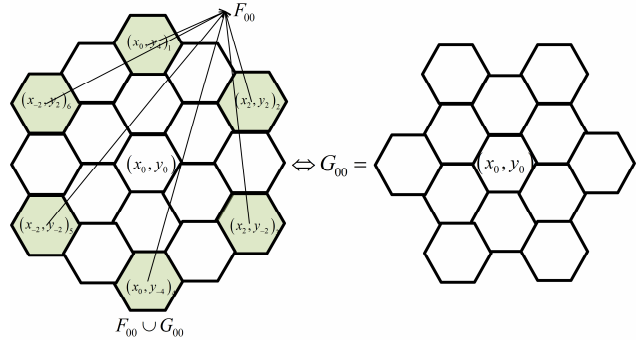

Fig.7 Graphic representation of $F_{00}$ and $G_{00}$ sets

Consequently, $\forall (i,j) / C_{i,j} \in C, C_{m,n} \in F_{ij}$

$$C_{m,n} = \begin{Bmatrix} C_{ij+4}, C_{i+2\,j+2}, C_{i+2\,j-2}, \\ C_{ij-4}, C_{i-2\,j-2}, C_{i-2\,j+2} \end{Bmatrix}$$

So,

$$\forall C_{i,j} \in C \text{ and } \forall C_{m,n} \in G_{ij} / (i,j) \neq (m,n) \Rightarrow CCH_{ij} \neq CCH_{mn} \quad (11)$$

$$\exists (C_{i,j}, C_{m,n}) \in C \; / C_{m,n} \notin G_{ij} \Rightarrow CCH_{ij} = CCH_{mn} \quad (12)$$

According to (11), (12) and Fig.8 we can work into a given set $G_{ij}$ then we generalize the result for the entire network $C$.

Taking for example the sub network $G_{24}$. Let $CCH_1$ be allocated to cell $C_{2,4}$, in this case $\forall C_{m,n} \in G_{24} / (m,n) \neq (2,4)$ must not use the channel $CCH_1$.

Let $G_{24}(V,E)$ be the graph of sub network $G_{24}$. The application of Zykov's algorithm to $G_{24}(V,E)$ produces nine sub graphs as the case of 12 cells. In final phase ($9^{th}$ step), we are left with a complete graph with 4 vertices. The optimal solution is given by the complete graph in which vertex $C_{2,4}$ is allocated to the first color, $C_{1,1}, C_{1,5}, C_{3,3}$ and $C_{3,7}$ to the second color $C_{2,2}, C_{2,6}, C_{0,4}$ and $C_{4,4}$ to the third color and vertices $C_{1,3}, C_{1,7}, C_{3,1}$ and $C_{3,5}$ to the fourth.

To generalize this result you can apply the same procedure to all sub networks $G_{ij} \subset C$.

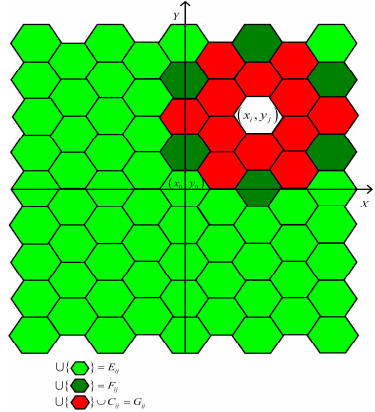

Fig. 8 Graphic representation of $E_{ij}$, $F_{ij}$ and $G_{ij}$ sets

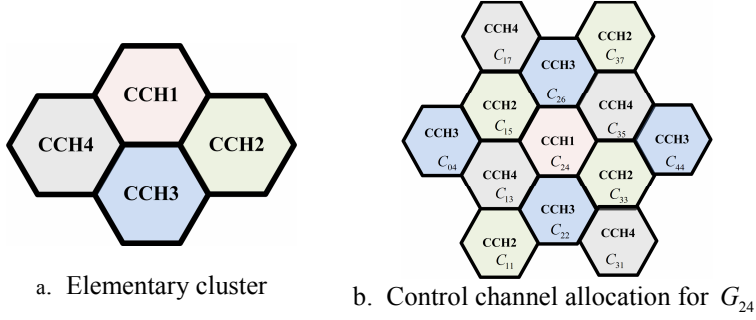

a. Elementary cluster  b. Control channel allocation for $G_{24}$

Fig. 9 Control channel allocation

**III - 1 - DATA CHANNELS**

*III - 1 - a - Case of network composed of 12 cells*

For data communication, we propose the use of the non overlapping channels (0, 1, 2, 3, 5, 6, 8, 9, 10, 12, 13 and 14) and the supplementary overlapping channels with their appropriate sequence code.

***Notations.***

- $N_c$ : the total number of cells with radius $R$,
- $D_{N_c x N_c}$ distance matrix, distance separating each couple of cell centres
- $N_{tch}$ : set of total channels,
- $N_{tdch}$ : set of total data communication channels,
- $N_{dch}$ : set of available data communication channels, it represents a sub set of total data communication channels set $N_{tdch}$.

$$N_{tdch} = N_{tch} - N_{cch}$$
$$N_{dch} \subseteq N_{tdch}$$

According to worldwide UWB regulations:
  o For US regulation : $N_{dch} = N_{tdch}$, $Card(N_{dch}) = Card(N_{tdch}) = 28$,
  o For European regulation : $N_{dch} \subset N_{tdch}$, $Card(N_{dch}) = 14$,
  o For Japan's regulation : $N_{dch} \subset N_{tdch}$, $Card(N_{dch}) = 20$,

For data communication, we assume that all PAN members transmit with the same transmitter power $P_t$, in order to have a coverage of radius $r$.

$$r < \frac{R}{2} \qquad (13)$$

In order to ensure efficient energy management, we propose a multi-hop routing inside each cell which is structured in mesh topology. One hop must be equal to or shorter than $r$ in order to:
- decrease transmit power to save sensor battery, maximize network life time and avoid interference,
- balance energy consumption and load over all cells of the network.

The choice of $P_t$ is done by taking into account the following equation.

$$P_{Rx} = P_t + Pl(r) / P_{Rx} - Link\_margin = Rx\_sensitivity \qquad (14)$$

As shown in Fig.11 taking the example of cell number 4, PAN members except PAN coordinator can be located at any position inside their cell.

So, we note for the case of sensors located at or near the cell border that they can interfere with sensors of adjacent PANs, located at or near their cell border.

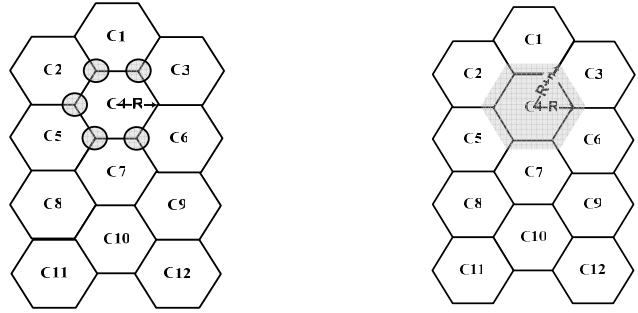

a. Radio coverage limit of a sensor     b. Radio coverage limit of a PAN

Fig.10 Radio coverage limit of a logical data communication channel

Consequently, as shown in Fig.10 the minimum distance of frequency reuse must be strictly bigger than $R_c$ (worst case). Let $D'_{min}$ represents the minimal distance of frequency reuse referring to cells centers or positions of PANs coordinators:

$$D'_{min} > R_c \qquad (15)$$

With $R_c$ is given by:

$$R_c = R + r \qquad (16)$$

From (15) and (16) and according to network hexagonal representation, the shortest distance frequency reuse at cells centers will be:

$$D'_{min} = 3R \qquad (17)$$

In this part, we are interested to find the minimal number of data communication channels to cover the global network taking into account frequency reuse. Similar to the case of control channel allocation, this problem can be translated into graph coloring problem applied to the graph shown by Fig.11a such that no two adjacent vertices share the same color. For that we can also call to Zykov's algorithm. The application of Zykov's algorithm to the previous graph produces ten sub graphs.

In step 10, we are left with a complete graph. The optimal solution is given by the complete graph (Fig.11 b) with three vertices. We note that to cover all the network from communication traffic without suffering from co-channel interference, we just need of 3 different channel frequencies (See Fig.12 a and Fig.12 b).

In the present case, the minimal number of data communications channels $N_{dch-opt}$, to cover the totality of the network by one data channel per PAN, is equal to three.

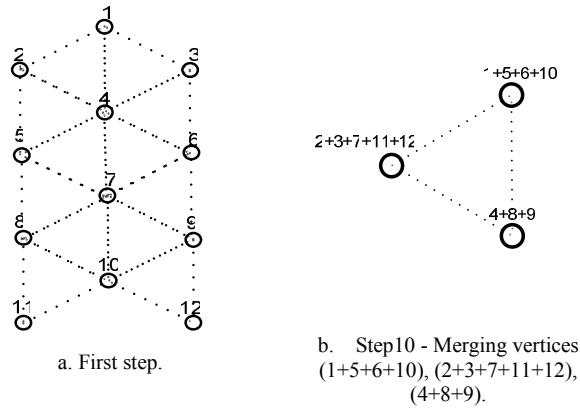

a. First step.     b. Step10 - Merging vertices (1+5+6+10), (2+3+7+11+12), (4+8+9).

Fig.11 Logical data communication channel allocation graph

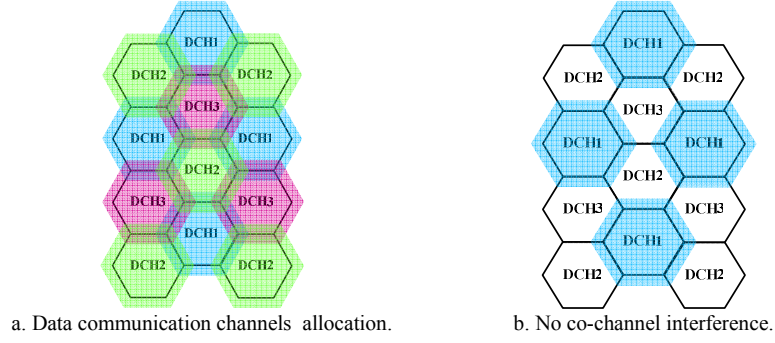

a. Data communication channels allocation.    b. No co-channel interference.

Fig.12 Logical data communication channel allocation

*III - 2 - b - General case*

**Theorem2**

Given a WSN composed by $N_c$ cells of radius $R$ organized on mesh topology with a node coverage equals to $r$, to totally cover the network by data traffic with one data channel per cell without suffering from co-channel interference, we need at most of 3 different channel frequencies.

**Proof2**

- Case Number of adjacent cells = 2:
- Case Number of adjacent cells = 3:

Let $C_{i,j}$, $C_{m,n}$ and $C_{k,l}$ be three adjacent cells (e.g. $d|C_{i,j}, C_{m,n}| = d|C_{i,j}, C_{k,l}| = d|C_{k,l}, C_{m,n}| < D_{min}$) $\Rightarrow$ the number of needed control channels is equal to 3,

- Case Number of adjacent cells $\geq 4$ :

We can distinguish the following cases:

- Case $N_c = 1$:

The number of needed data channel is equal to 1,

- Case Number of adjacent cells = 2:

Let $C_{i,j}$ and $C_{m,n}$ be two adjacent cells (e.g. $d|C_{i,j}, C_{m,n}| < D'_{min}$) $\Rightarrow$ the number of needed control channels is equal to 2,

- Case Number of adjacent cells $\geq 3$ :

For the case of data communication, similar to Proof1 done for control channel allocation (case of adjacent cells number $\geq 4$), to allow frequency reuse for a given couple of cells $\{C_{i,j}, C_{m,n}\} \subset C$, the distance separating the two cells centers must be equal to or bigger than $D'_{min}$.

$$d|C_{i,j}, C_{m,n}| \geq D'_{min}$$

$$\sqrt{(x_i - x_m)^2 + (y_j - y_n)^2} \geq 3R$$

$$(x_i - x_m)^2 + (y_j - y_n)^2 \geq 9R^2$$

$$\left(\left[x_0 + i \times \left(\frac{3R}{2}\right)\right] - \left[x_0 + m \times \left(\frac{3R}{2}\right)\right]\right)^2 + \left(\left[y_0 + j \times \left(\frac{\sqrt{3}R}{2}\right)\right] - \left[y_0 + n \times \left(\frac{\sqrt{3}R}{2}\right)\right]\right)^2 \geq 9R^2 \quad (18)$$

$$\left(\left(\frac{3R}{2}\right) \times (i - m)\right)^2 + \left(\left(\frac{\sqrt{3}R}{2}\right) \times (j - n)\right)^2 \geq 9R^2$$

$$3 \times (i - m)^2 + (j - n)^2 \geq 12$$

Let sub sets $E'_{ij}$, $F'_{ij}$ and $G'_{ij}$ given as follows:

$$E'_{ij} = \begin{cases} C_{m,n} \in C / \\ 3\times(i-m)^2 + (j-n)^2 > 12 \text{ and} \\ \text{fixed } C_{ij} \in C \end{cases}, \quad F'_{ij} = \begin{cases} C_{m,n} \in C / \\ 3\times(i-m)^2 + (j-n)^2 = 12 \text{ and} \\ \text{fixed } C_{ij} \in C \end{cases} \text{ and } G'_{ij} = C - \{E_{ij} \cup F_{ij}\}$$

You can rewrite $F'_{ij}$ as

$$3\times(i-m)^2 + (j-n)^2 = 12 \Rightarrow \begin{cases} \begin{cases}(i-m)^2 = 1, (j-n)^2 = 9 \Rightarrow |i-m|=1, |j-n|=3, \\ (i-m)^2 = 4, (j-n)^2 = 0 \Rightarrow |i-m|=2, |j-n|=0, \end{cases} \\ \Rightarrow \begin{cases} i = m+1, j = n+3 \\ i = m-1, j = n+3 \\ i = m+1, j = n-3 \\ i = m-1, j = n-3 \\ i = m+2, j = n = 0 \\ i = m-2, j = n = 0 \end{cases} \end{cases} \quad (19)$$

As done in Proof1, we conclude that $\forall (i,j)/C_{i,j} \in C$, $F'_{ij}$ and $G'_{ij}$ can be rewritten as follows:

$$F'_{ij} = \begin{cases} C_{i-1,j-3}, & C_{i+1,j-3}, & C_{i-1,j+3} \\ C_{i+1,j+3}, & C_{i-2,j}, & C_{i+2,j} \end{cases}$$

$$G'_{ij} = \begin{cases} C_{i,j}, & C_{i,j+2}, & C_{i,j-2} \\ C_{i+1,j+1}, & C_{i-1,j-1}, & C_{i-1,j+1} \\ C_{i+1,j-1} \end{cases}$$

So,

$$\forall C_{i,j} \in C \text{ and } \forall C_{m,n} \in G'_{ij}/(i,j) \neq (m,n) \Rightarrow DCH_{ij} \neq DCH_{mn} \quad (20)$$

$$\exists (C_{i,j}, C_{m,n}) \in C / C_{m,n} \notin G'_{ij} \Rightarrow DCH_{ij} = DCH_{mn} \quad (21)$$

According to (20) and (21), we can work into a fixed set $G'_{ij}$ then we generalize the result for the entire network $C$.

Taking for example the sub network $G'_{24}$, let $DCH_1$ be allocated to cell $C_{2,4}$, in this case $\forall C_{m,n} \in G'_{24}/(m,n) \neq (2,4)$ must not use the channel $DCH_1$.

Let $G'_{24}(V,E)$ be the graph of $G'_{24}$.

The application of Zykov's algorithm to the previous graph produces five sub graphs. In step 4, we are left with a complete graph with 3 vertices. The optimal solution is given by the complete graph in which vertex $C_{2,4}$ is allocated to the first color, $C_{1,3}, C_{2,6}$ and $C_{3,3}$ to the second color $C_{1,4}, C_{2,2}$ and $C_{3,5}$ to the third.

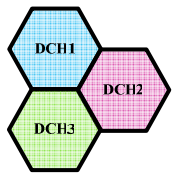   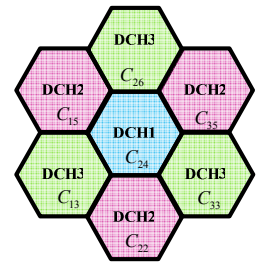

a. Elementary cluster          b. Control channel allocation inside $G'_{24}$.

Fig.13 Data channel allocation

To generalize this result you can apply the same procedure to all sub networks $G'_{ij} \subseteq C$.

## IV - DYNAMIC DATA CHANNELS ALLOCATION

## IV - 1 - General case of dynamic channels allocation

According to available channel frequencies $N_{dch}$ and PANs duty cycle, each PAN can benefit simultaneously from several data channels.
We define $K$ as :

$$K = ceil\left[ Card(N_{dch}) < Div\ N_{dch-opt} \right]$$

$K$ represents the number of simultaneous data channel frequencies that can benefit each PAN. From a spectrum regulation to another, $K$ isn't the same, for US: $K = 8$, Europe: $K = 4$ and Japan: $K = 6$.
But in reality each PAN is characterized by its duty cycle or superframe duration, as shown in Fig.14. So, according to PANs duty cycle $K$ can change during global network active period.
Let us assume the general case of a network composed by $N_c$ PAN coordinators with correspondent superframe durations $\{PAN_i = (SD_i, BI_i)\}_{1 \leq i \leq N_c}$. We define $\overline{BI}_{maj}$, $\overline{SD}_{min}$ and $U$ as respectively the major cycle, the elementary active cycle (ie elementary time unit) and the number of elementary active cycle per major cycle.

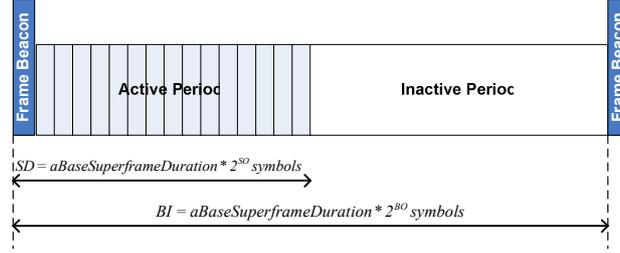

Fig.14 PAN superframe structure

$$\overline{BI}_{maj} = LCM(BI_1, BI_2, ...BI_{N_c}) = LCM(2^{BO_1}, 2^{BO_2}, ...2^{BO_{N_c}}) = \max_{1 \leq i \leq N_c}(2^{BO_i})$$

$$\overline{SD}_{min} = LCD(SD_1, SD_2, ...SD_{N_c}) = LCD(2^{SO_1}, 2^{SO_2}, ...2^{SO_{N_c}}) = \min_{1 \leq i \leq N_c}(2^{SO_i})$$

$$U = \frac{\overline{BI}_{maj}}{\overline{SD}_{min}}$$

Let $DC_{N_c xU}$ and $D_{N_c xN_c}$ represent respectively the matrix of duty cycle of all PANs coordinators per elementary active cycle and the matrix of distances separating each couple of cells center.

$$DC_{N_c xU} = \begin{bmatrix} DC_{11} & & DC_{1U} \\ DC_{21} & & \\ & & DC_{N_c-1U} \\ DC_{N_c 1} & & \end{bmatrix}$$

$$D_{N_c xN_c} = \begin{bmatrix} 0 & & D_{1N_c} \\ D_{21} & 0 & \\ & 0 & D_{N_c-1N_c-1} \\ D_{N_c 1} & & 0 \end{bmatrix}$$

So, given $DC_{N_c xU}$, $D'_{min}$ and $D_{N_c xN_c}$, we can determine the graph $G'(V, E)_i$ per elementary active cycle and then compute optimal data communication channels to cover the totality of the network as done in previous section (the case where all PAN coordinators are active).
In last step, we compute the number of simultaneous data communication channels $K_i$ per active PAN for the relative elementary active cycle.

$$K_i = ceil\left[ Card(N_{dch}) Div\ N_{dch-opt_i} \right]$$

In conclusion, considering the available data communication channels $N_{dch}$, $DC_{N_c xU}$, $D_{N_c xN_c}$, $D'_{min}$, we can compute the matrix of sub set of data communication channels per cell per elementary active cycle.

$$N_{dch_{N_c \times U}} = \begin{bmatrix} N_{dch_{11}} & & N_{dch_{1U}} \\ N_{dch_{21}} & & \\ & & N_{dch_{N_c-1U}} \\ N_{dch_{N_c 1}} & & \end{bmatrix}$$

For example $N_{dch_{11}}$ represents the set of data communication channels used by the first PAN (ID =1) during the first elementary active cycle, where $Card\left[N_{dch_{11}}\right] = K_1$.

With JAVA programming language (Eclipse-SDK-3.4.1) and MATLAB R2008a environment, proposed schemes are implemented. For static allocation our algorithm presents a complexity of order $O(n)$, where for dynamic allocation it is less fast and it presents a complexity of order $O(n^2)$.

**IV - 2 - Performance evaluation**

Let us consider a synchronized UWB-based WHSN of 12 PANs. Taking the example of the worst case given by Fig.15 where all PANs begin communication at the same time.
Let us assume that the European regulation is adopted (i.e. channels 4 and 7 for control and the rest for data).
Each $PAN_i$ is characterized by its superframe duration $(SD_i, BI_i)$ as shown by Fig.15.

So, $\overline{BI}_{maj} = 32$, $\overline{SD}_{min} = 1$, $U = \frac{\overline{BI}_{maj}}{\overline{SD}_{min}} = \frac{32}{1} = 32$.

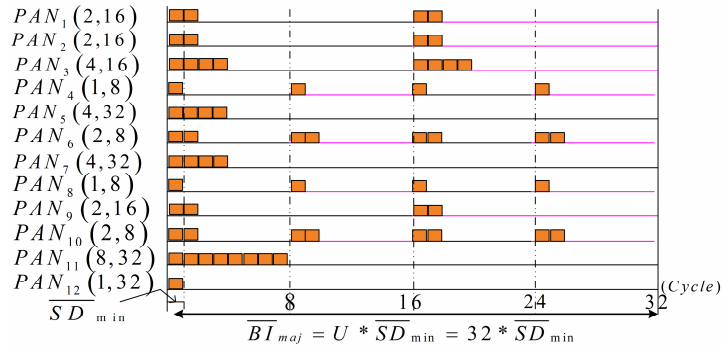

Fig.15 Example of PAN configuration

According to Fig.16 we note that:
- During the $1^{st}$, $2^{nd}$, $17^{th}$ and $18^{th}$ elementary cycles, each active PAN benefit simultaneously from 4 channels.
- During the $3^{rd}$, $4^{th}$, $9^{th}$ and $25^{th}$ elementary cycles each active PAN benefit from 7 channels (complete graph is composed by two vertices).
- During the $5^{th}$, $6^{th}$, $7^{th}$, $8^{th}$, $19^{th}$ and $20^{th}$ elementary cycles, only one PAN ($3^{rd}$ or $11^{th}$) is active which benefit simultaneously from all available channels.
- During the $10^{th}$ and $26^{th}$ elementary cycles only two PANs ($6^{th}$ and $10^{th}$) are active, each one benefit simultaneously from all available channels because the distance separating those two PANs is greater than $D'_{min}$.

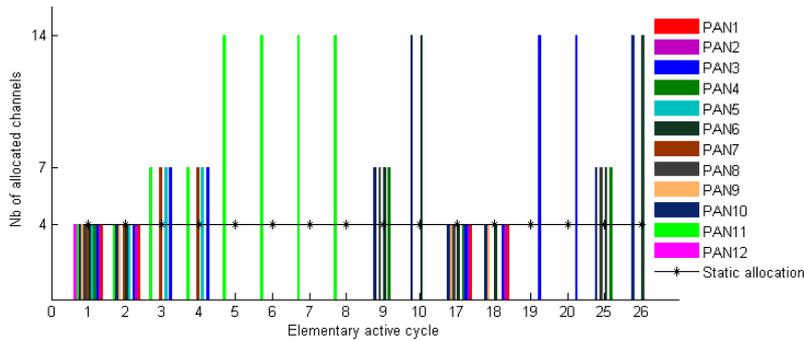

Fig.16 Data channels allocation during active elementary cycles

As illustrated in Fig.17, with static data channels allocation, the maximum number of allocated channels per PAN is 8, 6 and 4, respectively for US, Japanese and European regulation. Where with dynamic data channels allocation during specific elementary cycles active PANs can benefit from supplementary channels which are initially been allocated to some other PANs.

During the $5^{th}$, $6^{th}$, $7^{th}$, $8^{th}$, $10^{th}$, $19^{th}$, $20^{th}$ and $26^{th}$ active PANs benefit up to 28, 18 and 14 in respectively US, Japanese and European regulation.

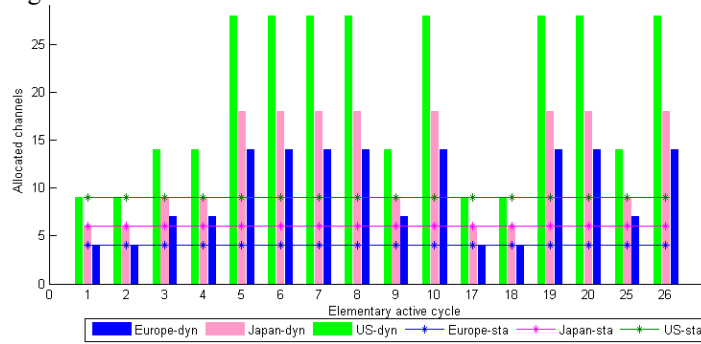

Fig.17 Static vs Dynamic channel allocation

Inside each active PANs and during each elementary cycle, let us assume the simple scenario of hight requests of three time slots each one. At the level of each active PAN coordinators, we suppose that requests are scheduled without any conflict.

According to Fig.18 and Fig.19 we note that:
- With single data channel and static multi-channel schemes, each active PAN needs respectively 24 and 6 time slots to answer to all requests. Although with static multi-channel scheme the results are extremely better than with single data channel scheme but we note a spectrum resource waste during PANs sleep period.

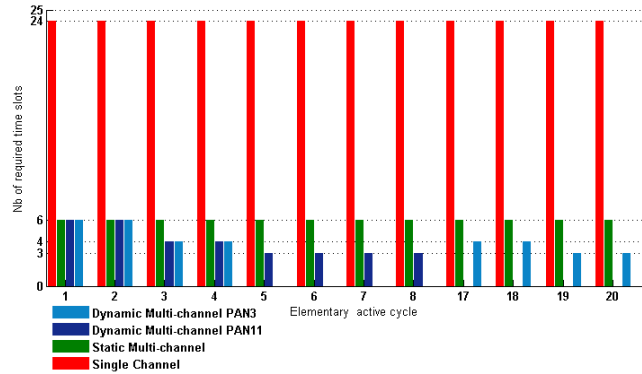

Fig.18 Required Time slots inside the $3^{rd}$ and $11^{th}$ PANs

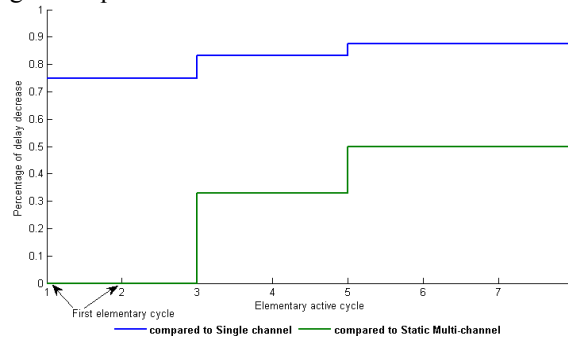

Fig.19 Percentage of delay decrease of the $11^{th}$ PAN

- With dynamic multi-channel scheme we note:

- o To answer to all requests, the $3^{rd}$ and $11^{th}$ PANs require only 4 time slots during $3^{rd}$ and $4^{th}$ elementary cycles and 3 time slots during respectively $(19^{th}, 20^{th})$ and $(5^{th}, 6^{th}, 7^{th}, 8^{th})$ elementary cycles.
- o In this way we can ensure, on the one hand, an efficient and fair data channels allocation between PANs permitting an enhancement of QoS inside each PAN and, on the other hand, a maximization of channel utility.

## V - CONCLUSION

As pronounced in the beginning of this paper, efficient allocation of the available spectrum resource in WSNs under scalable and optimal multi-frequency MAC protocols allowing parallel transmissions with optimal use of available resource, without suffering from interference, data communication conflict and control packet overhead, seem to be an imperative and challenging task.

To resolve such problem for large-scale and dense WSNs as WHSNs, we propose to decompose the frequency allocation problem into two sub-problems: static control channel allocation to ensure a permanent control frequency per PAN avoiding control channel congestion problem and dynamic data channel allocation based on PANs duty cycle information and spatial frequency reuse to avoid the underutilization of spectrum resource.

**Bibliography**
[1] X. Chen, P.H. Qiu-Sheng, H. Shi-liang, T.Zhang-Long, Chen , "A Multi-Channel MAC Protocol for Wireless Sensor Networks", The Sixth IEEE International Conference on Computer and Information Technology, Seoul, September 2006, pp. 224-224.
[2] X. Wang and T. Berger, "Spatial channel reuse in wireless sensor networks", Wireless Networks Journal, vol 14, iss 2, pp. 133-146, March 2008.
[3] CC2420 datasheet, 2004 chipcon,inst.eecs.berkeley.edu/cs150/Documents/CC2420.pdf
[4] BS. Jamila, S. Ye-Qiong, K. Anis, F. Mounir, "A Three-Tiered Architecture for Large-Scale Wireless Hospital Sensor Networks", the International Workshop on Mobilizing Health Information to Support Healthcare-Related Knowledge Work - MobiHealthInf 2009, pp 20-31.
[5] J. Jemai, R.Piesiewicz, T.Kurner, "Calibration of an indoor radio propagation prediction model at 2.4 GHz by measurements of the IEEE 802.11b preamble" IEEE $61^{st}$ Vehicular Technology Conference, Spring. 2005,Vol 1, pp.111-115.
[6] IEEE 802.15.4a Standard Part 15.4: IEEE Standard for Information Technology, Amendment to IEEE Std 802.15.4™-2006, 2007.
[7] A. Kathryn, "Classical Techniques", Springer US Book, chapter 2, 2005, pp.19-68.
[8] B. Raman, "Channel Allocation in 802.11-Based Mesh Networks", 25th IEEE International Conference on Computer Communications, INFOCOM 2006, pp. 1-10.
[9] A.H.M. Rad, V.W.S. Wong, "Joint channel allocation, interface assignment and MAC design for multi-channel wireless mesh networks", in Proceedings of IEEE INFOCOM, 2007, pp. 1469-1477.